# Unconventional spin dynamics in the non-collinear phase of a ferrimagnet

D.M. Krichevsky[1,2,3], N.A. Gusev[2,3], D.O. Ignatyeva[2,3,4], A.V. Prisyazhnyuk[3], E.Yu. Semuk[3], S.N. Polulyakh[3], V.N. Berzhansky[3], A.K. Zvezdin[2,5], V.I. Belotelov[2,3,4]

[1] Moscow Institute of Physics and Technology (MIPT), 141700 Dolgoprudny, Russia
[2] Russian Quantum Center, 121353 Moscow, Russia
[3] Physics and Technology Institute, Vernadsky Crimean Federal University, 295007 Simferopol, Russia
[4] Photonic and Quantum Technologies School, Lomonosov Moscow State University, 119991 Moscow, Russia
[5] Prokhorov General Physics Institute of the Russian Academy of Sciences, 119991, Moscow, Russia

**Ferrimagnets containing several partially compensated magnetic sublattices are considered the most promising materials for all-optical data storage and for ultrafast communications based on spin waves. There are two magnetic phases of the ferrimagnets: collinear and non-collinear ones. Up to now spin dynamics in ferrimagnets has been studied mostly in the collinear state without paying much attention to the kind of the magnetic phase. Here we investigate laser induced ultrafast spin dynamics in a rare-earth iron garnet film in the noncollinear phase as well. We identify a crucial influence of the magnetic phase on the excited spin modes which allowed us to discover several prominent effects previously overlooked. In particular, the non-collinearity makes the quasi-antiferromagnetic mode sensitive to the external magnetic field and brings its frequency close to the frequency of the quasi-ferromagnetic mode. The latter maximizes near the magnetization compensation point and vanishes towards the collinear phase. Spectacularly, at the phase transition the quasi-ferromagnetic mode becomes soft and its amplitude significantly increases reaching 7°. This opens new opportunities for the ultrafast control of spins in ferrimagnets for nonthermal data storage and data processing.**

1. ## INTRODUCTION

Ultrafast magnetic phenomena driven by femtosecond light pulses are emerging topics of modern material science. The growth in the number of studies in this field is stimulated by technological demands and fundamental puzzles ranging from ultrafast all-optical data storage[1–5] and spin transport [6,7] to quantum computing[8,9].

Ferrimagnets are unique among ordered magnetic materials as they combine properties of both ferro- and antiferromagnets. Being composed of several antiferromagnetically coupled magnetic ion sublattices, these materials sustain low frequency quasi-ferromagnetic (q-FM) and high frequency quasi-antiferromagnetic (q-AFM), so-called "exchange" modes[10]. The latter is not attainable by GHz conventional microwave stimuli and can be excited by femtosecond laser pulses due to either thermal or optomagnonic effects[11]. Ultrafast spin dynamics along with ultrafast switching, domain wall motion, spin waves, and skyrmion formation were extensively studied in ferrimagnetic metals, such as GdFeCo[5,12–16] and CoGd[17–19]. In contrast to metals and alloys, insulating oxides are almost lossless at optical frequencies. It makes them perfect materials for essential applications such as all-optical non-thermal magnetic recording[1,2] and control spin waves in all-dielectric nanostructures[20].

Another important feature of ferrimagnets is possibility for full compensation of their sublattices magnetic moments which takes place at some temperature $T_M$ called magnetization compensation point. The theory of magnetic phase transitions in compensated ferrimagnets was previously extensively studied[21–24]. In the H-T phase diagram of a ferrimagnet two main phases are possible: collinear and non-collinear ones. The collinear phase is accompanied with collinear alignment of magnetizations of both sublattices and external magnetic field. The non-collinear phase is characterized by bevel of sublattices magnetizations combined with a net magnetization slant to the external magnetic field.

Previous experiments on excitation of spin dynamics in rare-earth iron garnets (RIG) by femtosecond laser pulses were primarily concerned on the magnetic states in the collinear phase[10,25–28] and the kind of magnetic phase was not identified. An optical excitation of the exchange resonance between rare-earth and transition metal sublattices far from the compensation point was studied in[10] for an in-plane configuration of external magnetic field. The results were found to be in an agreement with the conventional Kaplan–Kittel theory[29,30] which describes q-AFM mode far from $T_M$. In [28,31] a comprehensive experimental study of RIG in an in-plane magnetic field was made in a wide temperatures range including magnetization compensation point, and both an exchange and ferromagnetic resonances were investigated. Nevertheless, magnetic phases were not identified and the features of spin dynamics were not observed. As a result, a simplified description of the mode frequencies behavior in terms of ferromagnetic resonance and Kaplan-Kittel theories satisfied the experimental data.

Importantly, the role of the magnetic phase on spin dynamics was not discussed previously and, in particular, the non-collinear phase was not addressed in this aspect. At the same time, nowadays, the room-temperature non-collinear phase, actively studied in various magnetic materials[32–34], is a potential candidate for chiral spin textures based magnetic non-volatile memory[35], spintronic devices[36] and even qubits for quantum computing. Hence, the understanding of spin dynamics in this phase is of urgent fundamental and practical demand.

In the current work we confront magnetic phase of a ferrimagnet with its ultrafast spin dynamics and highlight a crucial role of the ground state in the character of spin oscillations. Spin dynamics in the non-collinear phase of RIG in the vicinity of the magnetization compensation temperature point is studied comprehensively. We scrutinized magnetic phase transitions in the material via magneto-optical measurements and distinguished the temperature-dependent features of the spin modes by the magneto-optical pump-probe technique. The experimental results are well described by the quasi-antiferromagnetic theoretical approach.

## 2. RESULTS

Unique magnetic properties of RIG are governed by its complex crystal structure. $Fe^{3+}$ ions occupy sites in tetrahedral and octahedral sublattices of the cubic unit cell, with the total magnetization of the $Fe^{3+}$ ions in the tetrahedral positions being greater than in the octahedral ones. Uncompensated magnetic moment of $Fe^{3+}$ ions gives rise to net magnetization $\mathbf{M}_{Fe}$ which is antiparallelly ordered to magnetization $\mathbf{M}_R$ of rare-earth element occupying the third (dodecahedral) sublattice. Due to a huge exchange field between $Fe^{3+}$ ions in tetrahedral and octahedral sublattices they can be treated as a single one with magnetization $\mathbf{M}_{Fe}$ in this case. Spins of the rare-earth sublattice are ordered by an exchange field produced by iron ions. The existence of magnetic moment compensation point ($T_M$) is mainly due to a strong temperature dependance of $M_R(T)$, while $M_{Fe}$ is hardly affected[22].

The magnetization state of a RIG film can be defined using the sublattice magnetizations $\mathbf{M}_{Fe}$ and $\mathbf{M}_R$ forming Néel vector $\mathbf{L} = \mathbf{M}_{Fe} - \mathbf{M}_R$ and magnetization vector $\mathbf{M} = \mathbf{M}_{Fe} + \mathbf{M}_R$ (Fig.

1a). The angle $\theta = \frac{\theta_{Fe} - \theta_R}{2}$ characterizes the deviation of the Néel vector from the film plane, while the angle $\epsilon = \frac{\theta_{Fe} + \theta_R}{2}$ characterizes a bevel of the sublattices magnetization from a collinear antiparallel orientation. Here $\theta_{Fe}$ and $\theta_R$ are the angles between the film plane and $\mathbf{M}_{Fe}$ and $\mathbf{M}_R$, respectively. For vectors in the upper half-space of the x-z plane $\theta > 0$ and for the bottom half-space $\theta < 0$. Wherein, in statics $\mathbf{M}_{Fe,R}$ always lie in the plane formed by the magnetic field and anisotropy axis (Supplementary, Section II).

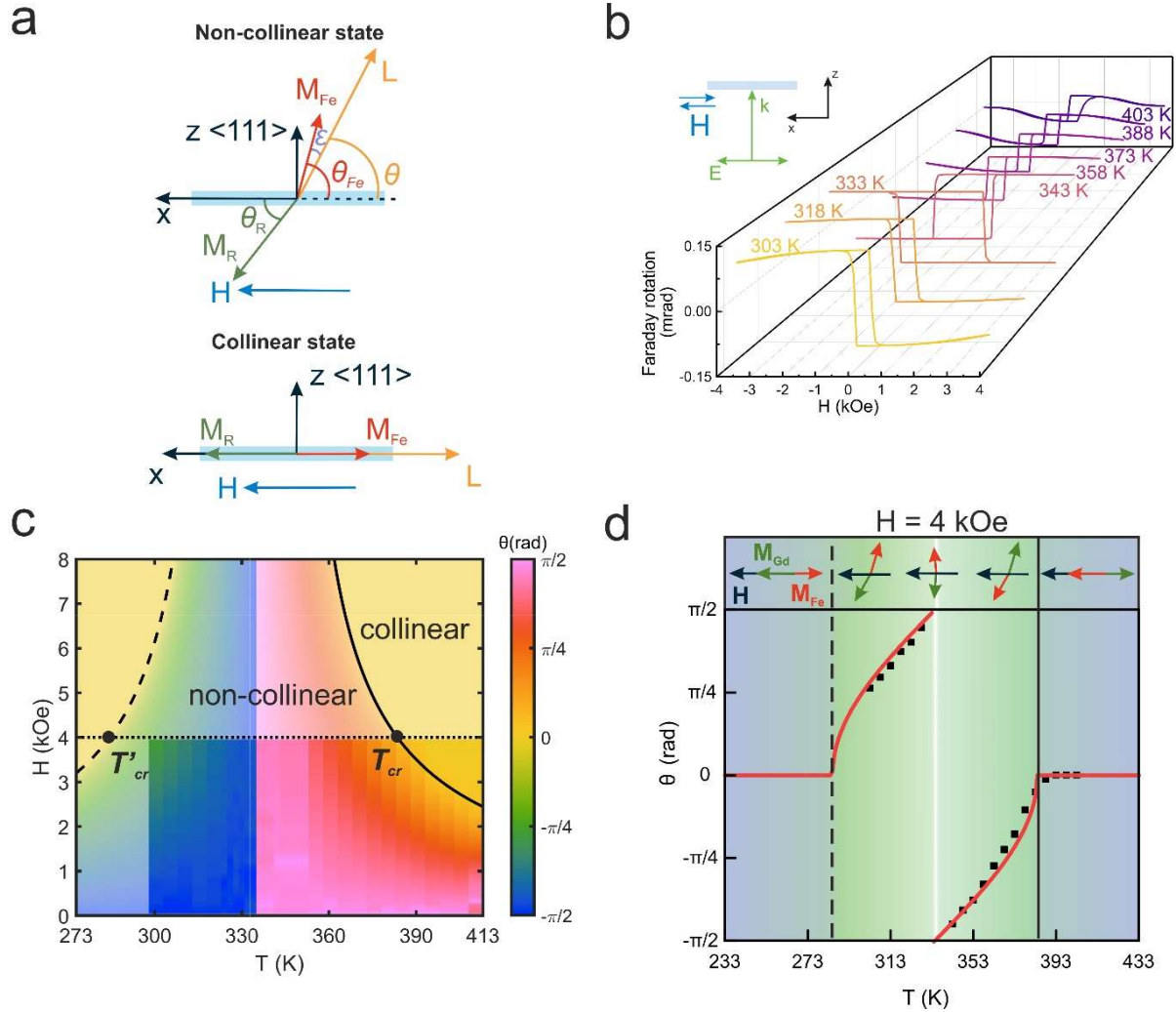

**Figure 1. (a) Configuration of the magnetic sublattices of the sample in the non-collinear and collinear phases. (b) The hysteresis loops for different temperatures of the sample (inset shows experimental configuration). (c) Experimental (bright) and theoretical (blurred) phase diagrams of the RIG film. Black curves (dashed and continuous) represent phase transition boundary between non-collinear and collinear phases. The dotted straight line represents the slice at 4 kOe made for (d). (d) Temperature dependence of $\theta$ for the applied in-plane 4 kOe magnetic field. Black points represent experimental data, whereas red line is theoretical approximation. The arrows show sublattices magnetization vectors orientation with respect to external magnetic field for collinear (blue filling), non-collinear (green filling) and compensated (white line) states. Vertical (dashed and continuous) lines represent boundaries between non-collinear and collinear phases.**

Here we investigate a RIG film with Gd rare-earth ions (see Methods). The film has a uniaxial magnetic anisotropy with the axis perpendicular to the film. We consider the case when the external magnetic field is in-plane and therefore perpendicular to the anisotropy axis. The analysis of the system's potential energy minimum reveals that in this case two magnetic phases are possible, depending on the temperature and external magnetic field: collinear and noncollinear (Supplementary, Section III). In the collinear phase the sublattice magnetic moments are antiparallel ($\varepsilon = 0$) and vectors **M** and **L** are aligned with the in-plane external magnetic field **H** ($\theta = 0$). The non-collinear phase provides the deflection of vectors **M** and **L** from **H** ($\theta \neq 0$). This deflection is accompanied with noncollinearity of the sublattices magnetic moment ($\varepsilon \neq 0$). The boarder between two phases is determined by the condition $m = m_{cr}$ (Supplementary, Section III). Here $m$ is a difference between sublattice magnetizations $m = M_{Fe}(T) - M_R(T)$, while $m_{cr}$ is a kind of critical magnetization defined by:

$$m_{cr} = \chi_\perp H + \frac{2K}{H}, \qquad (1)$$

where $\chi_\perp = \frac{(M_{Fe}(T)+M_R(T))^2}{2\Lambda M_{Fe}(T)M_R(T)}$ is magnetic susceptibility (Supplementary, Section III), $\Lambda$ is an exchange parameter between Fe and rare-earth sublattices, $K = K_{Fe} + K_R$, $K_{Fe}$ and $K_R$ are anisotropy constants of each sublattice (numerical values are given in Supplementary, Section I). The collinear phase exists for $|m| > m_{cr}$ and the non-collinear one establishes if $|m| < m_{cr}$. Thus, a second type phase transition from non-collinear to collinear phases occurs at $m = m_{cr}$. This transition is directly related to the critical temperature $T_{cr}$ at which $m = m_{cr}$.

To experimentally determine the magnetic phase diagram $\theta(T, H)$ we measured the magneto-optical hysteresis loops at various temperatures. Some of the loops are presented in Fig.1b. The experiment was carried out in an almost in-plane external magnetic field (the field's tilt angle was around 1 deg.) for the normal light incidence. Therefore, the measured Faraday rotation angle was proportional to the out-of-plane component of the sample gyration vector. Since the gyration in RIG is mostly provided by $Fe^{3+}$ ions[37], the observed Faraday rotation conforms to $\mathbf{M}_{Fe}$. Furthermore, since the tilt between Fe and Gd sublattices given by angle $\epsilon$ is small (Supplementary, Section II), such measurement provides information regarding deflection of the Néel vector ($\theta \approx \theta_{Fe}$). At zero magnetic field $\mathbf{M}_{Fe}$ is directed normally to the surface ($\theta_{Fe} = \pi/2$) due to the uniaxial magnetic anisotropy. A magnetic field applied in-plane deflects $\mathbf{M}_{Fe}$ from the normal. Consequently, $\theta$ is found as follows:

$$\theta(H,T) \approx \theta_{Fe}(H,T) = \mathrm{asin}\left(\frac{FR(H,T)}{|FR(0,T)|}\right), \qquad (2)$$

where $FR(H,T)$ and $FR(0,T)$ are angles of the Faraday rotation (FR) under applied magnetic field *H* and zero magnetic field, correspondingly.

The experimentally measured magnetic phase diagram $\theta(T, H)$ and its crossection at $H = 4$ kOe are given in Fig. 1c,d, respectively. A distinct jump of derivative $\partial\theta/\partial T$ takes place at $m = m_{cr}$ (solid black curve in Fig. 1c), e.g. at $T_{cr} \sim 383$ K for $H = 4$ kOe (Fig. 1d), indicating the second kind phase transition. The other phase transition of the second kind appears for smaller temperatures when $m < 0$ and $|m| = m_{cr}$ (dashed black curve in Fig. 1c) which corresponds to $T'_{cr} \sim 284$ K at H=4 kOe (Fig. 1d). For a fixed $H$ the noncollinear phase ($\theta \neq 0$) appears for the temperature interval $T'_{cr}(H) < T < T_{cr}(H)$. This temperature range decreases for larger magnetic fields (Fig 1c) which is due to the Zeeman energy of the sample. Notably, in the non-collinear phase theory predicts bistability of the system due the degeneracy of $+\theta$ and $-\theta$ states (see Supplementary, Section III, Fig. S3). However, in our experiments a small out-of-plane magnetic field component exists which lifts this degeneracy and selects the direction of $M_z$ projection parallel to this small out-of-plane field component. Hence **L** flips while crossing $T_M$ due to a change of sing of $m$ at the temperature $T_M$ of the magnetization compensation, which indicates that for the studied sample $T_M = 336$ K.

Ultrafast spin dynamics in the sample was excited by fs-laser pump pulses at 787 nm and observed by the Faraday rotation (FR) of the fs-laser probe pulses at 515 nm delayed with respect to the pump pulse in the configuration shown in Fig. 2a (Methods). The pump-probe experiment was conducted in a wide temperature range which allowed us to observe features of the non-collinear states below and above $T_M$. Moreover, we resolved the spin dynamics at the phase transition ($T_{cr}$=383 K at $H = 4$ kOe) and in the collinear phase, for larger temperatures (Fig.2b). Figure 2b,c demonstrate dynamics of the Néel vector in terms of the time-resolved Faraday polarization rotation (TRFR) at $H = 4$ kOe. Two kinds of oscillations are clearly observed: low frequency ones at the time scale up to 350 ps (Fig.2b) and high frequency ones at the time scale of 50 ps (Fig. 2b and zoomed view in Fig. 2c). Notably, the high-frequency component in the TRFR signal was not observed for the temperatures above $T_M$. Data on the frequency and magnitude of these modes were extracted from gauged TRFR signals and compared with the theoretical values (Fig.3 a,b).

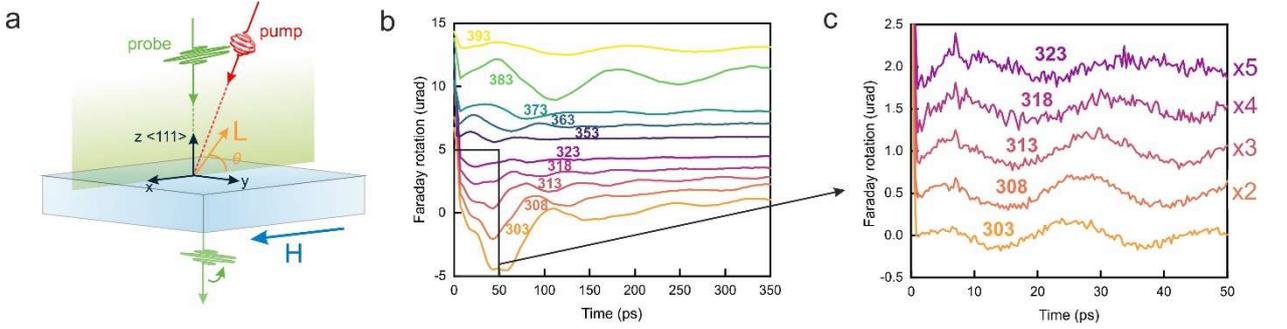

**Figure 2. Configuration of the pump-probe experiment (a). Ultrafast magnetization dynamics represented by the probe Faraday transients in the range of 350 ps for different temperatures from 303 K (below T$_M$) to 393 K (above T$_M$): (b) general view for the range up to 350 fs and (c) zoomed view for the range of 50 ps demonstrating the higher frequency mode in detail. The data in (c) are presented after subtracting the lower-frequency mode. The amplitude of the oscillations in (c) for 308-323 K are scaled (the scale factor is presented on the right side of the figure).**

Theoretical analysis of the spin dynamics in the current ferrimagnetic system is performed based on the Euler-Lagrange equations of motion (Supplementary, Section II). In our calculations we assumed that gyromagnetic ratio of Fe and Gd ions are equal ($\gamma_{Fe} = \gamma_{Gd} = \gamma$)[38]. In the case of non-collinear phase ($|m| < m_{cr}$ or $\epsilon \neq 0$ ) the mode frequencies are given by (Supplementary, Section IV):

$$\omega_{\text{q–AFM,q–FM}} = \left( \Omega_1^2 + \Omega_2^2 + \omega_0^2 \pm \sqrt{(\Omega_1^2 + \Omega_2^2)^2 + 2\omega_0^2 \Omega_2} \right)^{\frac{1}{2}}, \quad (3)$$

where $\Omega_1^2 = \frac{\omega_H^2}{2} - \omega_H \omega_{KK} \cos\theta$, $\Omega_2^2 = \frac{1}{2}(2\omega_H \cos\theta - \omega_{KK})^2$, $\omega_H = \gamma H$, $\omega_{KK} = \gamma \frac{|m|}{\chi_\perp}$. In Eq. (3) the sign "+" corresponds to the higher frequency (q-AFM) while the sign "-" corresponds to the lower frequency(q-FM) (inset in Fig. 3a). Noteworthy, the term "quasi" is typically used for canted

AFMs, such as FeBO₃, owing uncompensated magnetic moment[39] which is similar to the ferrimagnetic iron garnets. Frequencies of the modes for collinear state are presented in Supplementary, section IV. The frequencies calculated using Eq. (3) (solid curves in Fig.3a,b) agree well with the experimental findings (dots in Fig. 3a,b).

The features of the q-AFM and q-FM modes in the non-collinear state differ drastically from those in the collinear phase far from $T_M$. In the latter case, the q-AFM mode has the character of the Kaplan-Kittel exchange mode, whose frequency is unaffected by the external magnetic field and is primarily influenced by the exchange magnetic field [10,25,28]. However, in the non-collinear state, the situation dramatically changes: the q-AFM mode frequency becomes magnetic field-dependent (Fig.3a, brown symbols). This provides an important and convenient tool for its control that was previously unavailable in the collinear states far from $T_M$.

The behavior of the q-FM mode is nearly the opposite. Typically, the frequency of q-FM mode depends on the applied magnetic field in the collinear state (e.g. see ref. [28]). However, in the non-collinear state this turns upside-down: the q-FM mode becomes hardly sensitive to the magnetic field, as shown in Fig.3a by blue symbols.

Moreover, our study shows (Fig.3b) that near the compensation point $f_{q-AFM}$ and $f_{q-FM}$ become close to each other: $f_{q-AFM}$ decreases and $f_{q-FM}$ increases for the temperature increase towards $T_M$ so that both modes have extremums at this temperature. There is only a small frequency gap at $T_M$: $\Delta\omega = \gamma \left( \sqrt{\frac{2K}{\chi_\perp} + H^2} - \sqrt{\frac{2K}{\chi_\perp}} \right)$, which can be controlled by external magnetic field. This situation is also quite unusual since, conventionally, at the temperatures far from $T_M$, the frequencies of the two modes have several orders difference. For example, frequencies of 3 GHz and 410 GHz were observed for the q-FM and q-AFM modes for the films without compensation point at similar experimental conditions[25].

We broadened the view of modes frequencies behavior and calculated frequencies of q-FM and q-AFM modes for different values of the external magnetic field and temperature using Eq. 3 (Fig. 3 c-d). A boundary between the collinear and non-collinear phases are shown in Fig. 3c,d by dashed white line. A peculiar non-monotonous dependence of $f_{q-FM}$ on the magnetic field is observed in the non-collinear phase near the phase transition temperature $T_{cr}$: it decreases with the increase of the magnetic field (Fig.3c) which is in a high contrast with what happens in the collinear phase and in the ferromagnetic materials. Notably, the theory predicts that at $T_{cr}$ $f_{q-FM}$ tend to zero (Fig. 3a,c).

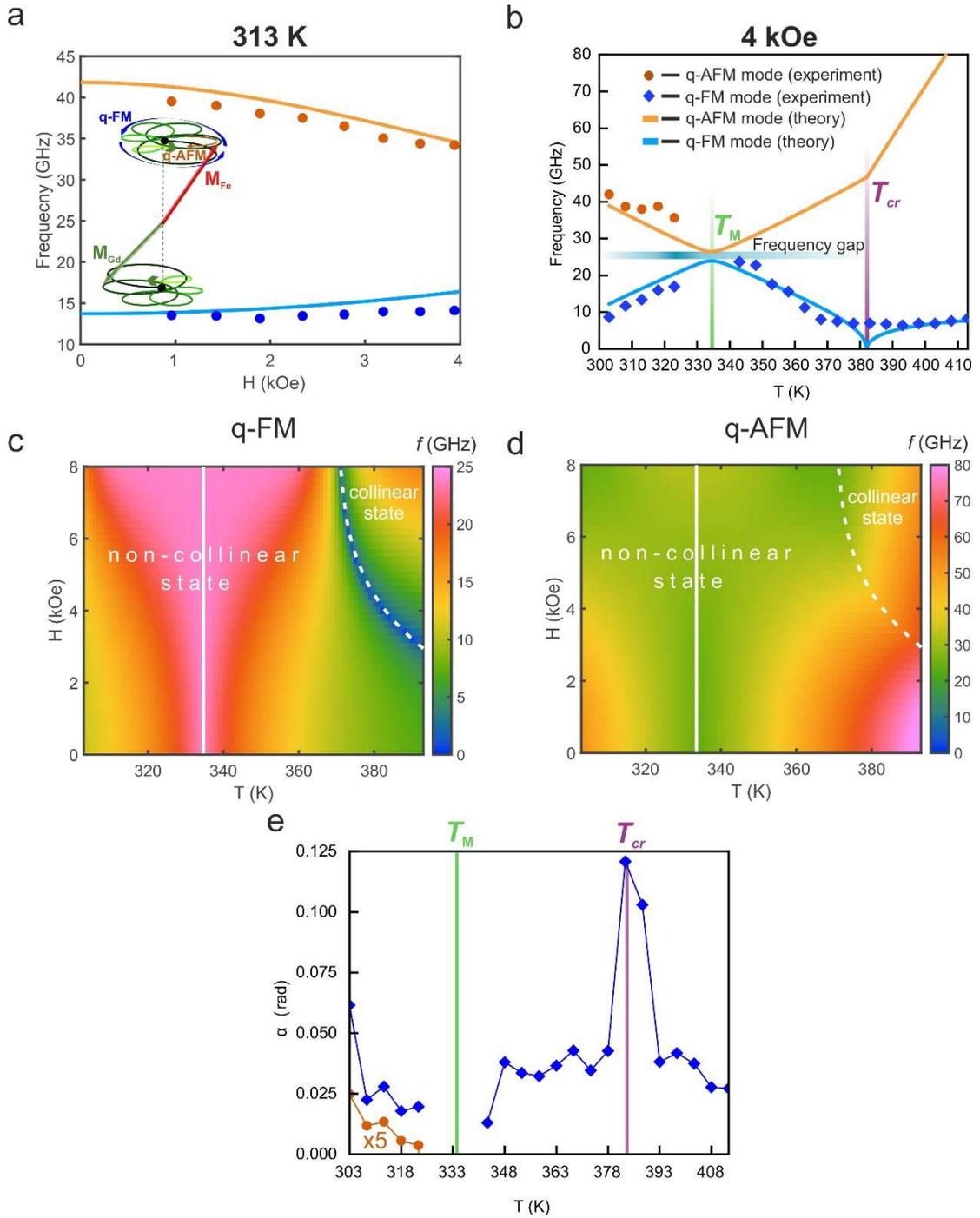

**Figure 3.** (a,b) Experimentally found frequencies of the q-FM (blue dots) and q-AFM (brown dots) modes compared to theoretical curves calculated from Eq.(3): (a) versus magnetic field at $T$=313 K and (b) versus temperature for $H = 4$ kOe. The inset in (a) shows complex trajectory of $M_{Fe}$ and $M_R$ vectors during laser-induced simultaneous excitation of the q-FM and q-AFM modes. The trajectory is shown for the time interval equal to one period of the q-FM mode. Six periods of the q-AFM mode are seen. Color of the trajectory curve denotes temporal coordinate, so that the dynamics starts at the black point and proceeds to the light green color. (c-d) Calculated oscillation frequencies of the q-AFM (c) and q-FM modes (d) as a function of external magnetic field and temperature. White dashed lines at (c) and (d) indicate the boundary of phase transition between non-collinear and collinear phases. (e) Amplitude of the spin dynamics in terms of the magnetization deflection angle from the ground state versus temperature for the q-FM (blue dots) and q-AFM modes (brown dots) at $H = 4$ kOe.

Let's now consider spin dynamics at around second kind phase transition between the non-colinear and colinear phases, where the frequency curve derivative $\partial f/\partial T$ breaks (at around $T_{\text{cr}}$ = 383 K, $H = 4$ kOe). At the boundary between the non-collinear and collinear phases, an intriguing feature appears. The q-FM mode's amplitude increases pronouncedly and reaches a maximum at $T_{\text{cr}}$ (Fig. 3e). At this point, its frequency drops down to 7 GHz (Fig. 3a), which indicates a soft mode character [24]. This phenomenon is explained by a significant increase of the magnetic susceptibility which takes place at the magnetic phase transition of the second kind. We should emphasize that here we demonstrate optical excitation of the soft mode in the ferrimagnetic dielectric at room temperature which is quite important for advanced ultrafast spin control. The excitation efficiency of spin dynamics at the soft mode is enhanced by around 4 times if compared to the collinear phase near the transition temperature (for instance, at 413 K) and by 10 times in comparison to the non-collinear phase (at 343 K).

### 3. CONCLUSION

To conclude, in this study, we identify an importance of the magnetic phase of a ferrimagnet for its ultrafast spin behavior. A rare-earth iron garnet near magnetization compensation temperature was considered. We demonstrated several crucial peculiarities of spin dynamics in a non-collinear state that contrast sharply with the usually observed spin dynamics of the exchange and ferromagnetic modes in a collinear state far from the compensation point. In particular, when temperature approaches the compensation point the frequencies of q-AFM and q-FM modes behave oppositely: the former decreases, while the latter one grows. The situation changes after crossing the compensation point for higher temperatures. We also discovered that transition from the non-collinear phase to the collinear one is accompanied with softening of the q-FM mode which leads to a huge increase of the excitation efficiency and amplitude. The amplitude of the soft mode becomes more than 4 times larger than for the collinear state (at 413 K) and up to 10 times higher than for the non-collinear phase (at 343 K). As the deflection angle of the soft mode was found to reach ~7°, it can be potentially interesting for nonlinear magnonic phenomena such as Bose-Einstein condensation[40,41] and superfluidity as well as for all-optical switching.

We also found that, in contrast to the conventional case of the collinear phase where the q-AFM mode is field-independent and q-FM mode strongly depends on field. In the noncolinear phase the behavior becomes upside-down: q-AFM mode turns to a field-dependent character, while the q-FM mode gets almost field independent which is in agreement with our theoretical predictions based on q-AFM approximation of Euler-Lagrange equations of motion. The described methodology allows temperature control of magnetization states of RIG for magnonic and spintronic devices. The approach described in current study is universal and can be applied for RIG with various rare-earth ions as well as for other ferrimagnets with uniaxial anisotropy.

### 4. MATERIALS AND METHODS

*A. Sample*

The sample is a 2.2 µm thick RIG film of composition $(Bi_{0.6}Gd_{2.4})(Fe_{4.28}Ga_{0.57}Ge_{0.15})O_{12}$ where rare-earth ions is gadolinium (Gd). The film is grown on (111) $(CaGd)_3(GaZrMg)_5O_{12}$ substrate and has a uniaxial magnetic anisotropy with the axis perpendicular to the film.

The $Bi^{3+}$ ions have a large ionic radius, therefore, for the synthesis of Bi substituted RIG films the $(CaGd)_3(GaZrMg)_5O_{12}$ substrate with the large lattice parameter $a_s$ = 1.2494 nm was chosen. The RIG film was grown by liquid phase epitaxy method from an overcooled solution-melt on a horizontally fixed (111) substrate at the isothermal conditions. The $Bi_2O_3$–$B_2O_3$ -PbO oxides were used as a solvent. Sample growth temperature $T_g$ was 742.5 C.

A mismatch between the crystal lattice of the substrate and the magnetic film, $\Delta a$= -0.007 A was determined by the X-ray diffraction method. The film thickness $h$=2.2 µm were found from optical transmittance spectra. A magnetopolarimeter was used to measure the magnitude of the Faraday effect wavelength of 515 nm 4.54 deg/µm. The compensation temperature $T_M$=336K was determined by the sign change of the Faraday effect.

*B. Static and time-resolved magneto-optical measurements*

For static magneto-optical (MO) measurements of hysteresis loops the sample was placed into in-plane external magnetic field of electromagnet (AMT&C Troitsk). The beam with ~250 fs pulse duration was produced from a tunable optical parametric amplifier (Avesta PARUS), which was pumped by a 1 kHz high-energy Yb regenerative amplifier (Avesta TETA). Linearly polarized (E vector directed along external magnetic field) normally incident 515 nm light pulses were utilized for polarization rotation measurements. The light was modulated by optical chopper (Thorlabs) at 500 Hz. Passed through the sample light pulses were detected using balanced photodetector (Newport Nirvana 2007). Electrical signal from photodetector was detected using lock-in detector (Zurich instruments MFLI).

For time-resolved magneto-optical measurements the sample was placed into constant in-plane external magnetic field of electromagnet (AMT&C Troitsk). 787 nm circularly polarized pulses pumped the sample at ~10° of polar angle. Linearly polarized along external magnetic field probe pulses of 515 nm hit the sample at normal incidence. The pump and probe pulses were focused on the sample to a 100 µm and 40 µm spots correspondingly. The pump and probe beams fluence were 30 $mJ/cm^2$ and 0.3 $mJ/cm^2$ correspondingly. Temporal overlap between pump and probe pulses was varied by using a 600 mm motorized translation stage (Thorlabs) with a retroreflector in the control beam optical path. The pump pulses were modulated by optical chopper (Thorlabs) at 500 Hz. Passed through the sample probe pulses were detected using balanced photodetector (Newport Nirvana 2007). Electrical signal from photodetector was detected using lock-in detector (Zurich instruments MFLI). The sample was heated by Peltier element electrically controlled with current stabilization. Temperature of the sample was controlled by thermistor.

*C. Theoretical description*

The RIG ferrimagnetic film was theoretically analyzed using a two-sublattice model. Quasi-antiferromagnetic approximation was used to describe the magnetization behavior near the compensation point, which means the two sublattice magnetizations were considered nearly antiparallel expect for a small bevel angle. To obtain the static ground state of the ferrimagnet, we calculated minima of the potential energy using Lagrange and Hamiltonian functions. Dynamics of the system was analyzed using Euler-Lagrange equations of motion. The presented approach can be used for a wide class of the ferrimagnetic materials, therefore we put a thorough and detailed description of this analysis to Supplementary, see Sections II-IV.

## ACKNOWLEDGEMENTS

This work was financially supported by the Ministry of Science and Higher Education of the Russian Federation, Megagrant project N 075-15-2022-1108. We thank N.E. Khokhlov for help in preparation of the experimental set-up for the pump-probe experiments.

## REFERENCES

1. Stupakiewicz, A., Szerenos, K., Afanasiev, D., Kirilyuk, A. & Kimel, A. V. Ultrafast nonthermal photo-magnetic recording in a transparent medium. *Nature* **542**, 71–74 (2017).

2. Stupakiewicz, A. *et al.* Ultrafast phononic switching of magnetization. *Nat Phys* **17**, 489–492 (2021).

3. Kimel, A. V. & Li, M. Writing magnetic memory with ultrashort light pulses. *Nat Rev Mater* **4**, 189–200 (2019).

4. Gorchon, J. *et al.* Single shot ultrafast all optical magnetization switching of ferromagnetic Co/Pt multilayers. *Appl Phys Lett* **111**, 042401 (2017).

5. Stanciu, C. D. *et al.* Subpicosecond magnetization reversal across ferrimagnetic compensation points. *Phys Rev Lett* **99**, 14–17 (2007).

6. Hortensius, J. R. *et al.* Coherent spin-wave transport in an antiferromagnet. *Nat Phys* **17**, 1001–1006 (2021).

7. Kampfrath, T. *et al.* Terahertz spin current pulses controlled by magnetic heterostructures. *Nat Nanotechnol* **8**, 256–260 (2013).

8. Zou, J. *et al.* Domain wall qubits on magnetic racetracks. (2022).

9. Yuan, H. Y., Cao, Y., Kamra, A., Duine, R. A. & Yan, P. Quantum magnonics: When magnon spintronics meets quantum information science. *Phys Rep* **965**, 1–74 (2022).

10. Parchenko, S. *et al.* Non-thermal optical excitation of terahertz-spin precession in a magneto-optical insulator. *Appl Phys Lett* **108**, 1–5 (2016).

11. Kirilyuk, A., Kimel, A. V. & Rasing, T. Ultrafast optical manipulation of magnetic order. *Rev Mod Phys* **82**, 2731–2784 (2010).

12. Radu, I. *et al.* Transient ferromagnetic-like state mediating ultrafast reversal of antiferromagnetically coupled spins. *Nature* **472**, 205–209 (2011).

13. Becker, J. *et al.* Ultrafast Magnetism of a Ferrimagnet across the Spin-Flop Transition in High Magnetic Fields. *Phys Rev Lett* **118**, 1–5 (2017).

14. Vahaplar, K. *et al.* Ultrafast Path for Optical Magnetization Reversal via a Strongly Nonequilibrium State. *Phys Rev Lett* **103**, 66–69 (2009).

15. Stanciu, C. D. *et al.* Ultrafast spin dynamics across compensation points in ferrimagnetic GdFeCo: The role of angular momentum compensation. *Phys Rev B Condens Matter Mater Phys* **73**, 1–4 (2006).

16. Vahaplar, K. *et al.* All-optical magnetization reversal by circularly polarized laser pulses: Experiment and multiscale modeling. *Phys Rev B Condens Matter Mater Phys* **85**, 1–17 (2012).

17. Binder, M. *et al.* Magnetization dynamics of the ferrimagnet CoGd near the compensation of magnetization and angular momentum. *Phys Rev B Condens Matter Mater Phys* **74**, 1–5 (2006).

18. Caretta, L. *et al.* Fast current-driven domain walls and small skyrmions in a compensated ferrimagnet. *Nat Nanotechnol* **13**, 1154–1160 (2018).

19. Kim, K. J. *et al.* Fast domain wall motion in the vicinity of the angular momentum compensation temperature of ferrimagnets. *Nat Mater* **16**, 1187–1192 (2017).

20. Chernov, A. I. *et al.* All-Dielectric Nanophotonics Enables Tunable Excitation of the Exchange Spin Waves. *Nano Lett* **20**, 5259–5266 (2020).

21. Davydova, M. D., Zvezdin, K. A., Kimel, A. V & Zvezdin, A. K. Ultrafast spin dynamics in ferrimagnets with compensation point. *Journal of Physics: Condensed Matter* **32**, 01LT01 (2020).

22. Davydova, M. D., Zvezdin, K. A., Becker, J., Kimel, A. V. & Zvezdin, A. K. H-T phase diagram of rare-earth–transition-metal alloys in the vicinity of the compensation point. *Phys Rev B* **100**, 064409 (2019).

23. Sabdenov, Ch. K. *et al.* Magnetic-field induced phase transitions in intermetallic rare-earth ferrimagnets with a compensation point. *Low Temperature Physics* **43**, 551–558 (2017).

24. AK Zvezdin & AF Popkov. MAGNETIC-RESONANCE IN FERROMAGNETS WITH COMPENSATION POINT. *SOLID STATE PHYSICS* **16**, 1082–1089 (1974).

25. Parchenko, S., Stupakiewicz, A., Yoshimine, I., Satoh, T. & Maziewski, A. Wide frequencies range of spin excitations in a rare-earth Bi-doped iron garnet with a giant Faraday rotation. *Appl Phys Lett* **103**, 1–5 (2013).

26. Reid, A. H. M., Kimel, A. V., Kirilyuk, A., Gregg, J. F. & Rasing, Th. Optical Excitation of a Forbidden Magnetic Resonance Mode in a Doped Lutetium-Iron-Garnet Film via the Inverse Faraday Effect. *Phys Rev Lett* **105**, 107402 (2010).

27. Deb, M., Molho, P., Barbara, B. & Bigot, J. Y. Controlling laser-induced magnetization reversal dynamics in a rare-earth iron garnet across the magnetization compensation point. *Phys Rev B* **97**, 1–6 (2018).

28. Deb, M., Molho, P., Barbara, B. & Bigot, J. Y. Temperature and magnetic field dependence of rare-earth↔iron exchange resonance mode in a magnetic oxide studied with femtosecond magneto-optical Kerr effect. *Phys Rev B* **94**, 1–5 (2016).

29. Kaplan, J. & Kittel, C. Exchange Frequency Electron Spin Resonance in Ferrites. *J Chem Phys* **21**, 760–761 (1953).

30. Kittel, C. On the Theory of Ferromagnetic Resonance Absorption. *Physical Review* **73**, 155–161 (1948).

31. Deb, M., Molho, P. & Barbara, B. Magnetic damping of ferromagnetic and exchange resonance modes in a ferrimagnetic insulator. *Phys Rev B* **105**, 014432 (2022).

32. Sahoo, R. *et al.* Compensated Ferrimagnetic Tetragonal Heusler Thin Films for Antiferromagnetic Spintronics. *Advanced Materials* **28**, 8499–8504 (2016).

33. Kumar, V. *et al.* Detection of antiskyrmions by topological Hall effect in Heusler compounds. *Phys Rev B* **101**, 014424 (2020).

34. Choi, W.-Y., Yoo, W. & Jung, M.-H. Emergence of the topological Hall effect in a tetragonal compensated ferrimagnet $Mn_{2.3}Pd_{0.7}Ga$. *NPG Asia Mater* **13**, 79 (2021).

35. Hirata, Y. *et al.* Vanishing skyrmion Hall effect at the angular momentum compensation temperature of a ferrimagnet. *Nat Nanotechnol* **14**, 232–236 (2019).

36. Céspedes-Berrocal, D. *et al.* Current-Induced Spin Torques on Single GdFeCo Magnetic Layers. *Advanced Materials* **33**, 2007047 (2021).

37. Zvezdin, A. K., Kotov, V. A. *Modern magnetooptics and magnetooptical materials.* (CRC Press, 1997).

38. Drovosekov, A. B. *et al.* Magnetization and ferromagnetic resonance in a Fe/Gd multilayer: experiment and modelling. *Journal of Physics: Condensed Matter* **29**, 115802 (2017).

39. Mashkovich, E. A. *et al.* Terahertz Optomagnetism: Nonlinear THz Excitation of GHz Spin Waves in Antiferromagnetic $FeBO_3$. *Phys Rev Lett* **123**, 157202 (2019).

40. Bunkov, Yu. M. *et al.* Quantum paradigm of the foldover magnetic resonance. *Sci Rep* **11**, 7673 (2021).

41. Giamarchi, T., Rüegg, C. & Tchernyshyov, O. Bose–Einstein condensation in magnetic insulators. *Nat Phys* **4**, 198–204 (2008).

# Supplemental Material

# Ultrafast spin dynamics near magnetization compensation point in the non-collinear state of rare-earth garnets

D.M. Krichevsky[1,2,3], N.A. Gusev[2,3], D.O. Ignatyeva[2,3,4], A.V. Prisyazhnyuk[3], E.Yu. Semuk[3], S.N. Polulyakh[3], V.N. Berzhansky[3], A.K. Zvezdin[2,5], V.I. Belotelov[2,3,4]

[1] Moscow Institute of Physics and Technology (MIPT), 141700 Dolgoprudny, Russia
[2] Russian Quantum Center, 121353 Moscow, Russia
[3] Physics and Technology Institute, Vernadsky Crimean Federal University, 295007 Simferopol, Russia
[4] Photonic and Quantum Technologies School, Lomonosov Moscow State University, 119991 Moscow, Russia
[5] Prokhorov General Physics Institute of the Russian Academy of Sciences, 119991, Moscow, Russia

Contents
Section I. Determination of the magnetic characteristics of the sample
Section II. The Lagrangian of a two-sublattice system in the quasi-antiferromagnetic approximation
Section III. The potential energy and equilibrium states of the system
Section IV. Spin dynamics and mode frequencies

## Section I. Determination of the magnetic characteristics of the sample

The comparison of theory with experimental data becomes valuable only when one uses the information of the magnetic values of the sample, needed for calculations, such as the saturation magnetization $M_s$, uniaxial anisotropy constant $K$, etc. For the proposed study, it is necessary to know the value of the saturation magnetization depending on the temperature $M_s(T)$ in a range in the vicinity of the compensation temperature point. Although one can easily find temperature curve $M_s$ of iron-garnet [1,2], it may vary depending on the composition of the material. For this reason we calculate a temperature magnetization curve using simulation molecular field theory (Fig. S1). The curve corresponds to the composition $(Bi_{0.6}Gd_{2.4})(Fe_{4.28}Ga_{0.57}Ge_{0.15})O_{12}$, which is used in our experimental studies, is presented in Figure S1. As it can be seen, the compensation temperature is 336 K, which is in full agreement with our experimental results. At the same time, in the vicinity of the compensation point, the magnetization lies in the range of 0–5 emu/cm$^3$, which is also in agreement with the experimental data from the literature for the samples of other composition [1,2]. The dependence in Fig. S1 allows one to match the theoretical and experimental curves due to connection between the temperature $T$ (which is recorded in the experimental study) and the magnetization $M$ (which the theory operates on).

Uniaxial anisotropy constant $K$ and the parameter $\chi_\perp$ (see below) were estimated as 13500 erg/cm$^3$ and $3.7 \cdot 10^{-4}$ respectively on the base of the experimental data presented in the paper.

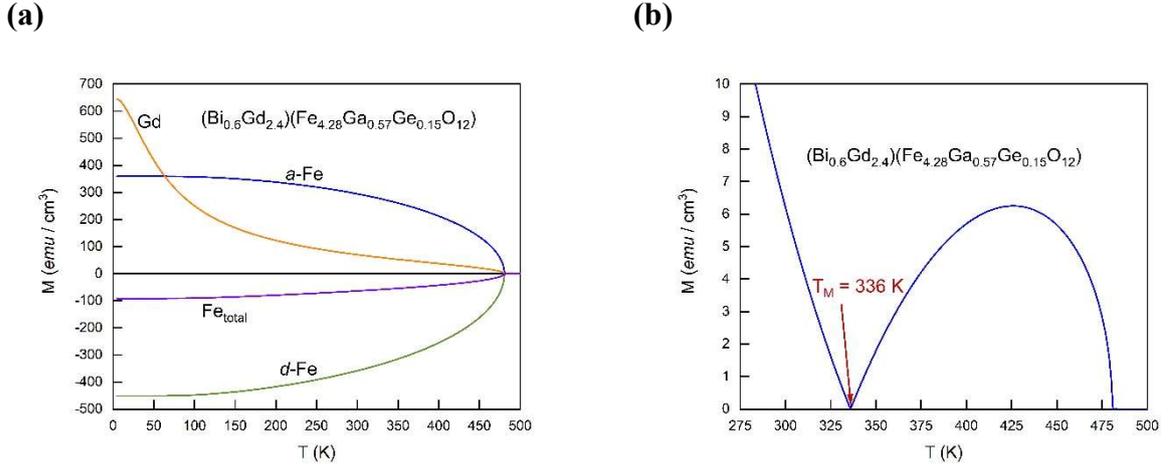

**Figure S1.** Temperature dependence of the sublattices magnetization (a) and net magnetization (b)

### Section II. The Lagrangian of a two-sublattice system in the quasi-antiferromagnetic approximation

The ferrimagnetic rear earth iron garnet (like the one in our studies) can be represented theoretically as the two-sublattice system. One of the sublattices is represented by Fe ions, and the other one - by rare-earth Gd ions which are coupled by the exchange interaction. The external magnetic field applied to the system can also be taken into account. Here we use the assumption that the applied external field is much smaller than the Hilbert exchange field.

Lagrangian method can be applied to describe the magnetization vectors of sublattices for obtaining the precession equations (oscillation equations or) of the antiferromagnetic vector.

The two-sublattice system is characterized by $\mathbf{M}_R$ and $\mathbf{M}_{Fe}$ magnetization vectors corresponding to the rare-earth and iron sublattices, respectively, and having $M_R$ and $M_{Fe}$ values and $\theta_R$, $\varphi_R$, and $\theta_{Fe}$, $\varphi_{Fe}$, angles in the spherical coordinate system (see Fig.1 in the main text). Using the well-known relation for the kinetic energy in micromagnetism [3–5], the Lagrange function $\mathcal{L}$ of the system can be written as

$$\mathcal{L} = -\frac{M_{Fe}}{\gamma_{Fe}} \sin\theta_{Fe} \frac{\partial \varphi_{Fe}}{\partial t} - \frac{M_R}{\gamma_R} \sin\theta_R \frac{\partial \varphi_R}{\partial t} - \Phi, \qquad (1)$$

where $\gamma_R$ and $\gamma_{Fe}$ are the gyromagnetic ratios, $\Phi$ is the potential energy, and $t$ is time. Lagrangian Eq. (1) is a function of 3 spatial coordinates and time, therefore a general analytical solution cannot be obtained in many cases. For the two-sublattice system, a so-called quasi-antiferromagnetic approximation considering nearly anti-parallel alignment of the sublattices can be used to simplify the description [3–5].

Therefore, antiferromagnetic vector $\mathbf{L} = \mathbf{M}_{Fe} - \mathbf{M}_R$ can be introduced and characterized by the angle set $\theta$ and $\varphi$, and:

$$\begin{aligned} \theta_{Fe} &= \theta + \varepsilon, \quad \theta_R = \varepsilon - \theta, \\ \varphi_{Fe} &= \varphi + \beta, \quad \varphi_R = \pi + \varphi - \beta \end{aligned} \qquad (2)$$

where the parameters $\varepsilon \ll 1$ and $\beta \ll 1$ characterize the noncollinearity of the magnetization vectors of the sublattices. If $\varepsilon = \beta = 0$, according to Eq. (2) the $\theta_{Fe} = -\theta_R$ and $\varphi_R = \pi + \varphi_{Fe}$, and that means the sublattice magnetization vectors are collinear and antiparallel. Therefore, for sublattice magnetization vectors one can obtain the following coordinates in Cartesian system:

$$\mathbf{M}_{Fe} = M_{Fe} \begin{cases} (\cos\theta - \varepsilon\sin\theta)(\cos\varphi - \beta\sin\varphi) \\ (\cos\theta - \varepsilon\sin\theta)(\sin\varphi + \beta\cos\varphi) \\ (\sin\theta + \varepsilon\cos\theta) \end{cases}$$

$$\mathbf{M}_R = M_R \begin{cases} (\cos\theta + \varepsilon\sin\theta)(-\cos\varphi - \beta\sin\varphi) \\ (\cos\theta + \varepsilon\sin\theta)(\beta\cos\varphi - \sin\varphi) \\ (\varepsilon\cos\theta - \sin\theta) \end{cases} \quad (3)$$

By substituting the expressions for the sublattice magnetization vectors Eq. (3) into Eq. (1), neglecting the terms $\varepsilon \cdot \dot{\beta}$ due to the smallness, and taking into account the property of Lagrangian function to be invariant under the addition of the full derivative of an arbitrary function of coordinates and time, one can obtain the Lagrangian of the two-sublattice system as a function of the angles $\theta$ and $\varphi$, parameters $\varepsilon$ and $\beta$, and their time derivatives:

$$\mathcal{L} = -\frac{m}{\gamma}\dot{\varphi}\sin\theta - \frac{\mathcal{M}}{\gamma}(\varepsilon\dot{\varphi} - \beta\dot{\theta})\cos\theta - \Phi, \quad (4)$$

Where $\mathcal{M}$ is the sum of the magnetization moduli of the sublattices, $\mathcal{M} = M_{Fe} + M_R$, the $\gamma = \gamma_R = \gamma_{Fe}$ is the gyromagnetic ratio and the $m = M_{Fe} - M_R$ is the difference between the magnetization moduli of the sublattices. Lagrangian function determined by Eq. (4) depends on the potential energy $\Phi$, which is determined by the material properties (the type of exchange interaction and magnetic anisotropy, shape and size that determine demagnetization energy, etc.) and the configuration of the applied magnetic field.

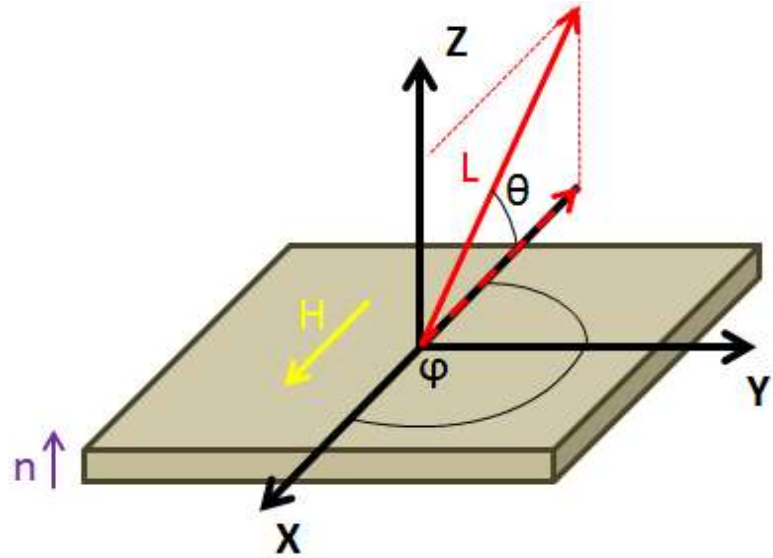

**Figure S2.** A ferrimagnetic film with uniaxial magnetic anisotropy in the external magnetic field **H**: the easy axis is denoted by the unit vector **n**, **L** is the aniferromagnetic vector. The $\theta$ angle is measured from the projection of the vector L onto the XOY plane.

We choose the coordinate system in the following way. XY plane coincides with the film plane; Z-axis is collinear to the film normal (Fig. S2). Uniform stationary external magnetic field **H** is applied along the X axis (**H**={*H*,0,0}, here and below we assume that *H*>0). The easy axis of the uniaxial effective magnetic anisotropy **n**={0,0,1} is directed along the film normal away from the substrate (the Z axis). The demagnetization energy for the thin film is proportional to $2\pi\left((\mathbf{M_{Fe}}+\mathbf{M_R}),\mathbf{n}\right)^2$ and is much smaller than the Zeeman and anisotropy energies in the vicinity of the compensation point, therefore it could be neglected. Thus, the potential energy of such system can be written as:

$$\Phi = -(\mathbf{M_{Fe}}+\mathbf{M_R})\mathbf{H} + \Lambda \mathbf{M_{Fe}}\mathbf{M_R} - K_{Fe}\frac{(\mathbf{M_{Fe}n})^2}{M_{Fe}^2} - K_R\frac{(\mathbf{M_R n})^2}{M_R^2} \qquad (5)$$

In Eq. (5) $K_{Fe}>0$ and $K_R>0$ are the magnetic anisotropy constants of each sublattice, $\Lambda>0$ is the intersublattice exchange constant and the "+" sign of the second term represents the antiferromagnetic character of the exchange interaction between the sublattices. Using Eq. (3) and taking into account **H** and **n** directions, one can rewrite (5) as:

$$\Phi = -mH\cos\theta\cos\varphi + \mathcal{M}H\beta\cos\theta\sin\varphi + \mathcal{M}H\varepsilon\sin\theta\cos\varphi + \frac{\delta}{2}(\varepsilon^2 + \beta^2\cos^2\theta - 1) - K\sin^2\theta \quad (6)$$

In Eq. (6) $\delta = 2\Lambda M_{Fe}M_R$, $K$ is the effective magnetic anisotropy constant, $K=K_{Fe}+K_R$. Considering $\varepsilon$ and $\beta$ as $O(1/\Lambda)$ the terms proportional to higher than $1/\Lambda$ powers are neglected to obtain Eq. (6). Thus, terms proportional to $\varepsilon^2 \cdot \beta^2$ were neglected for the exchange term in (6), the terms proportional to $\varepsilon \cdot \beta$ were neglected for the Zeeman term, and all terms proportional to $\varepsilon$ or $\beta$ were neglected for the anisotropy energy term.

Next, substituting the relation for the potential energy (6) into the Lagrange function (4) and considering ($\varepsilon,\beta$) as a generalized coordinates set, with respect to them one can calculate the Lagrange equations:

$$\frac{d}{dt}\frac{\partial \mathcal{L}}{\partial \dot{\varepsilon}} - \frac{\partial \mathcal{L}}{\partial \varepsilon} = 0$$
$$\frac{d}{dt}\frac{\partial \mathcal{L}}{\partial \dot{\beta}} - \frac{\partial \mathcal{L}}{\partial \beta} = 0$$

which gives the relation between two sets of coordinates, ($\varepsilon,\beta$) and ($\varphi,\theta$):

$$\varepsilon = -\frac{\mathcal{M}}{\delta}\frac{\dot{\varphi}}{\gamma}\cos\theta - \frac{\mathcal{M}H}{\delta}\sin\theta\cos\varphi$$
$$\beta\cos\theta = \frac{\mathcal{M}}{\delta}\left(\frac{\dot{\theta}}{\gamma} - H\sin\varphi\right) \qquad (7)$$

Substituting Eq. (7) into Eq. (4) with the potential energy Eq. (5), one can exclude the noncollinearity parameters $\varepsilon$ and $\beta$ and reduce the Lagrangian of the two-sublattice system to a function of only two angles $\varphi$ and $\theta$ of the antiferromagnetic vector **L** and their time derivatives:

$$\mathcal{L}_{eff} = \frac{\chi_\perp}{2}\left[\left(\frac{\dot{\varphi}}{\gamma}\cos\theta + H\sin\theta\cos\varphi\right)^2 + \left(H\sin\varphi - \frac{\dot{\theta}}{\gamma}\right)^2\right] - \dot{\varphi}\frac{m}{\gamma}\sin\theta + \\ + mH\cos\theta\cos\varphi + K\sin^2\theta \qquad (8)$$

where $\chi_\perp = \mathcal{M}^2/\delta$. Lagrangian in form Eq. (8), in contrast to Eq. (4), is the function of two variables, which greatly simplifies the system analysis.

**Section III. The potential energy and equilibrium states of the system**

The equilibrium states of the two-sublattice system are found as the minimum of the effective potential energy obtained using the known relationship between the Lagrange and the Hamiltonian functions for the stationary case $\dot{\varphi}=0$ and $\dot{\theta}=0$:

$$U_{eff} = -\frac{\chi_\perp}{2}H^2\sin^2\theta\cos^2\varphi - \frac{\chi_\perp}{2}H^2\sin^2\varphi - mH\cos\theta\cos\varphi - K\sin^2\theta \qquad (9)$$

Eq. (9) shows that antiferromagnetic vector **L** in the equilibrium state characterized by $\theta_0$, $\varphi_0$ angles always lie in the XZ plane ($\varphi_0=0$ for $m>0$ or $\varphi_0 = \pi$ for $m<0$). This can be explained using the first term in Eq. (5) that provides minima if the scalar product (**M**,**H**) of the ferromagnetic vector **M**=**M**$_{Fe}$+**M**$_R$ and external magnetic field **H** is positive: (**M**,**H**)>0. On the other hand, **L** and **M** vectors are nearly parallel for $m>0$ and nearly antiparallel for $m<0$.

The values of $\theta_0$ corresponding to the equilibrium state depend on the system parameters, as shown in Table 1.

**Table S1.** Equilibrium states of a two-sublattice magnetic system in the external field

|  | System parameters | Stationary value of angle $\theta$ | Stationary value of angle $\varphi$ | Scheme of the equilibrium state |
|---|---|---|---|---|
| I | $mH \geq \chi_\perp H^2 + 2K$ | $\theta_0 = 0$ | $\varphi_0 = 0$ | |
| II | $0 < mH < \chi_\perp H^2 + 2K$ | $-\pi/2 < \theta_0 < \pi/2$ $\cos\theta_0 = \dfrac{mH}{\chi_\perp H^2 + 2K}$ | | |
| III | $m = 0$ | $\theta_0 = \pm\pi/2$ | Not defined | |
| IV | $0 < |m|H < \chi_\perp H^2 + 2K$, $m < 0$ | $-\pi/2 < \theta_0 < \pi/2$, $\cos\theta_0 = \dfrac{-mH}{\chi_\perp H^2 + 2K}$ | $\varphi_0 = \pi$ | |

| V | $\|m\|H \geq \chi_\perp H^2 + 2K$, $m<0$ | $\theta_0 = 0$ | | 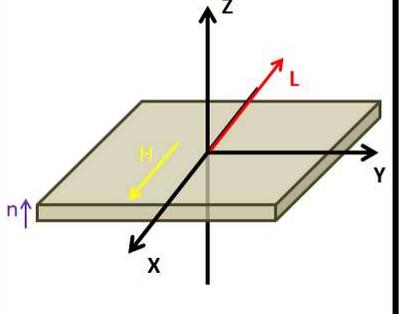 |

Cases (I) and (V), the so-called collinear phase, corresponding to $|m|H > \chi_\perp H^2 + 2K$ with $\theta_0 = 0$ describe antiferromagnetic vector **L** collinear to **H**. This is similar to the ferromagnetic state, as the ferromagnetic vector **M**=**M**$_{Fe}$+**M**$_R$ is parallel to **H**, while the parallel or antiparallel alignments of **L** and **H** are determined by the mutual orientation of **L** and **M** vectors that depends on sign of $m$.

Cases (II) and (IV), the so-called non-collinear phase, appear for rather small values of $m$ less than critical one $|m| < m_c = \chi_\perp H + 2K/H$. The direction of **L** in the equilibrium state is determined as:

$$\cos\theta_0 = \frac{|m|H}{\chi_\perp H^2 + 2K}. \quad (10)$$

It is important, that Eq. (10) has two solutions, so it means that there are two equilibrium states with $+\theta_0$ and $-\theta_0$. Moreover, Eq.(9) shows that these states are degenerate as they have the same potential energy $U_{eff}$. Therefore, for the same temperature that determines $m$, and the same external magnetic field there are two equivalent equilibrium positions described by $+\theta_0$ and $-\theta_0$ angles.

In the third case (III) describing the magnetization compensation point $m=0$, the antiferromagnetic vector **L** in the equilibrium state is directed strictly along the easy axis of the effective anisotropy **n**, and perpendicular to the external field **H**, so $\theta_0 = \pm\pi/2$.

Using Eq. (10), one can clarify the role of the $\chi_\perp$. By the calculation of the dependence **M**$_{Fe}$+**M**$_R$ = **M**(**H**), which can be performed using formulas (3), (7) and (10), one can easy confirm that the $\chi_\perp$ equals the magnetic susceptibility of the system at $m=0$ for the case **H** $\perp$ easy axis: $M_x = \chi_\perp H$.

In addition, according to Eq. (10) for $|m|=m_c$ $\theta_0=0$, and this is the second-order phase transition from the collinear to non-collinear state at this point. The magnetic susceptibility tensor element $\chi_{zx} = dM_z/dH_x$ equals zero for the collinear phase ($|m|>m_c$) and in the non-collinear one ($|m|<m_c$) is

$$\chi_{zx} = \pm\left(\chi_\perp\left(1-3\sin^2\theta_0\right) - \left(\frac{|m|}{H} + \frac{2\chi_\perp^2 H}{|m|}\right)\cos\theta_0\right)\cot\theta_0.$$

Thus, at the point $|m|=m_c$ the system exhibits the transition characterized by zeroing of the one of the susceptibility tensor element, so that it is similar to the transition from the ferromagnetic to the paramagnetic state at the Curie point. The magnetic susceptibility tensor element $\chi_{zx} = dM_z/dH_x$ may be treated as the order parameter of the system.

Figure S3 shows how these cases appear in the experimentally studied $(Bi_{0.6}Gd_{2.4})(Fe_{4.28}Ga_{0.57}Ge_{0.15})O_{12}$ film with the parameters, discussed in Section I. The transition between the collinear and non-collinear phases determined as $|m| = \chi_\perp H + 2K/H$ is shown by the black line. The value of $m \approx M_s$ (see Section I) since in the non-collinear phase the angle between $\mathbf{M_{Fe}}$ and $\mathbf{M_R}$ is less than 0.03 rad (see Fig. S4), and in the collinear state it equals 0. Therefore, $|\mathbf{M_{Fe}}-\mathbf{M_R}| \approx |\mathbf{M_{Fe}}|-|\mathbf{M_R}|=m$.

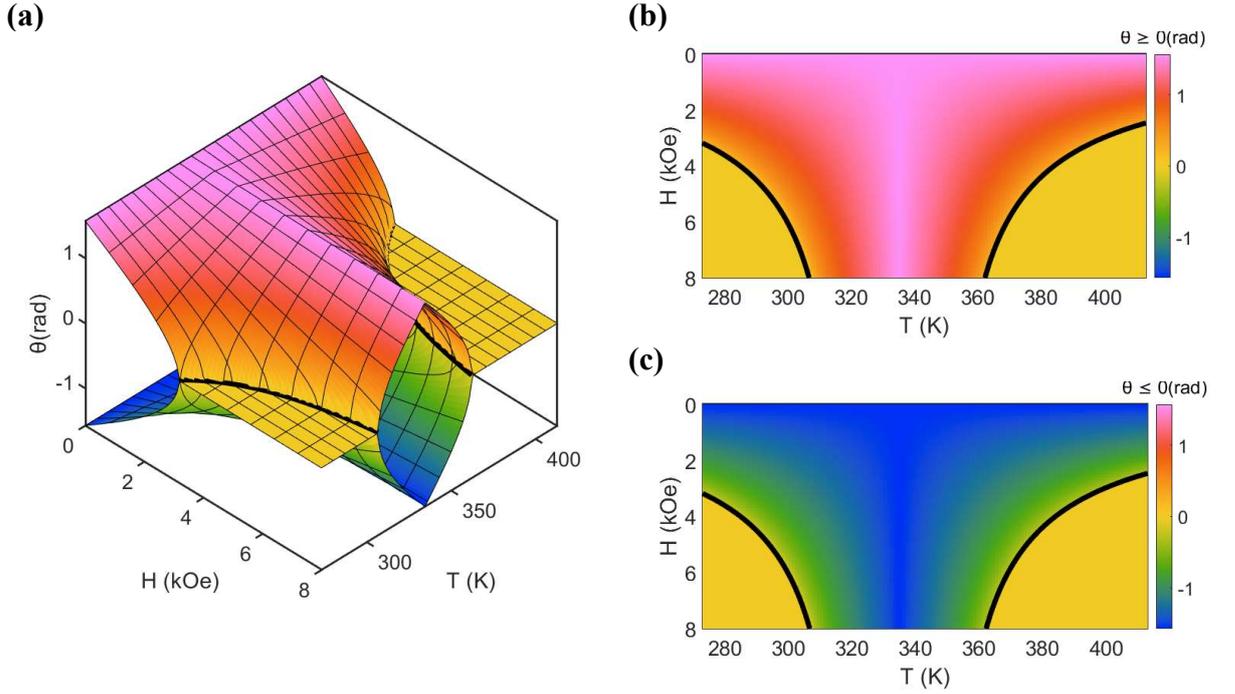

**Figure S3.** $\theta(T,H)$ diagrams showing (I)-(V) cases. (a) $\theta(T,H)$ 3D plot showing the degeneracy and non-collinearity in the vicinity of the magnetization compensation point. (b), (c) false-color plots showing $\theta(T,H)$ dependencies for $\theta \geq 0$ and $\theta \leq 0$ solution branches. $m=m_c$ condition is shown by the black line.

The cases (II), (III) and (IV) are accompanied by the physical phenomenon of static non-collinearity, which has a dual meaning. On the one hand, the antiferromagnetic vector $\mathbf{L}$ is non-collinear to the external magnetic field vector $\mathbf{H}$, as described above, and, on the other hand, the magnetization vectors of the $\mathbf{M_R}$ and $\mathbf{M_{Fe}}$ of the sublattices are also non-collinear to each other. It is convenient to discuss in terms of the bevel angle $\Delta$ between $\mathbf{M_{Fe}}$ and $-\mathbf{M_R}$: $\Delta = |\theta_R + \theta_{Fe}|$. Using Eq. (2), Eq. (7), and Eq. (10) $\Delta$ is expressed as:

$$\Delta = 2|\varepsilon| = \frac{2\mathcal{M}H}{\delta}|\sin\theta_0| \qquad (11)$$

The angle Δ reaches its maximum Δ=2$\mathcal{M}H$/δ at the magnetization compensation point $m$=0 (Fig. S4). With the increase of $m$, Δ decreases reaching Δ=0 at $|m|=m_c$. For $|m|\geq m_c$ where $\theta_0 = 0$ sublattice magnetizations are collinear to each other, and the case can be called the «collinear phase». At the same time, in both the collinear and non-collinear states $\mathbf{M_R}$ and $\mathbf{M_{Fe}}$ lie in the XZ plane and $\beta = 0$.

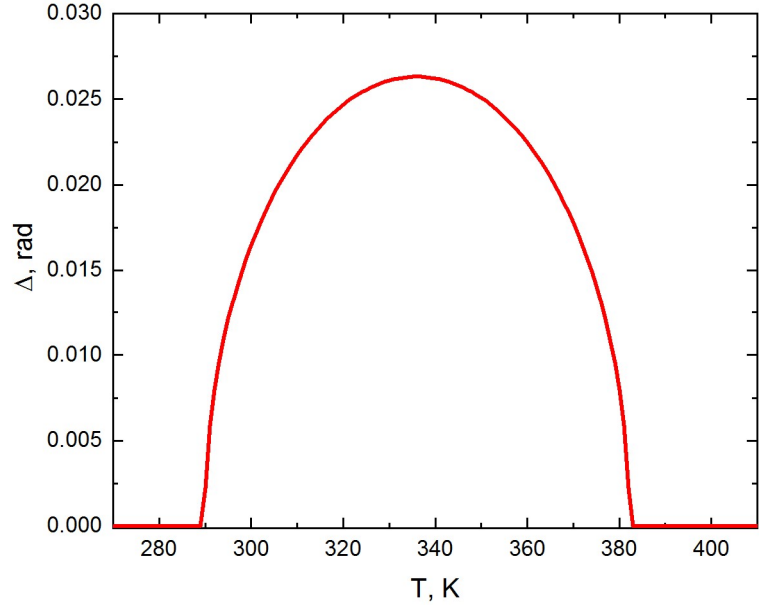

**Figure S4.** The Δ angle (in radians) vs temperature, according to Eq. (11).

We have considered the configuration where the external magnetic field **H** is applied strictly in-plane so that $\theta > 0$ and $\theta < 0$ states have the same energy according to Eq.(9). However, practically it is very complicated to apply the magnetic field strictly in-plane with a good precision, so that small out-of-plane magnetic field component always exists. The rigorous theoretical analysis of such a configuration is out of the scope of this manuscript. However, it is obvious that the presence of $H_z$ component lifts the degeneracy between the $\theta > 0$ and $\theta < 0$ states with $+M_z$ and $-M_z$ projections. According to Eq. (5), the state with $M_z H_z > 0$ has lower energy than with $M_z H_z < 0$. Thus, for the configurations where $H_z$ is not zero, the state with $M_z$ component parallel to this small out-of-plane field is preferable. As **L**=$\mathbf{M_R}$-$\mathbf{M_{Fe}}$ is nearly parallel to $\mathbf{M_R}$, $L_z$ and $M_z$ have the same signs for $m$>0 and the opposite signs for $m$<0. This results in the L flip in a vicinity of the compensation point due to a change of the balance between almost oppositely directed $\mathbf{M_{Fe}}$ and $\mathbf{M_R}$ vectors, i.e. to a change of sing of $m$ while crossing the compensation point (Fig. S5).

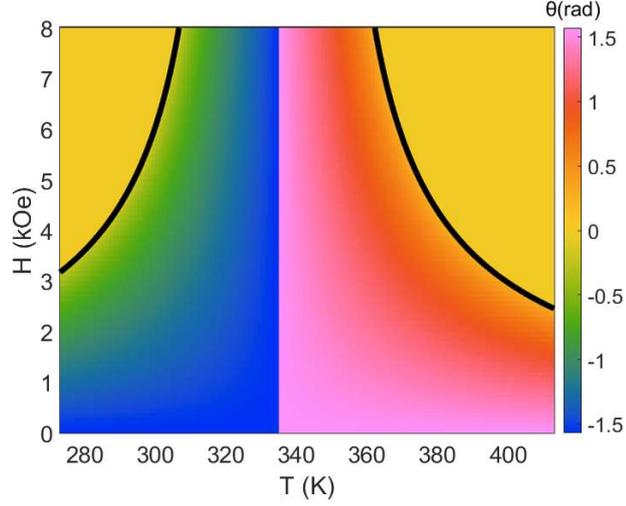

**Figure S5.** $\theta(H, T)$ diagram in the presence of the small out-of-plane magnetic field.

### Section IV. Spin dynamics and mode frequencies

In order to calculate the Lagrange equations for the angles $\theta$ and $\varphi$ of the antiferromagnetic vector **L**

$$\frac{d}{dt}\left(\frac{\partial \mathcal{L}_{\text{eff}}}{\partial \dot{\theta}}\right) - \frac{\partial \mathcal{L}_{\text{eff}}}{\partial \theta} + \frac{\partial \mathcal{R}}{\partial \dot{\theta}} = 0$$

$$\frac{d}{dt}\left(\frac{\partial \mathcal{L}_{\text{eff}}}{\partial \dot{\varphi}}\right) - \frac{\partial \mathcal{L}_{\text{eff}}}{\partial \varphi} + \frac{\partial \mathcal{R}}{\partial \dot{\varphi}} = 0$$

it is necessary to know the form of the Rayleigh function of the micromagnetic system. For the two-sublattice case, it can be written as follows [3–5]:

$$\mathcal{R} = \frac{\alpha}{2}\left(\frac{M_R}{\gamma_R}\left(\dot{\theta}_R^2 + \cos^2\theta_R \dot{\varphi}_R^2\right) + \frac{M_{Fe}}{\gamma_{Fe}}\left(\dot{\theta}_{Fe}^2 + \cos^2\theta_{Fe} \dot{\varphi}_{Fe}^2\right)\right) \quad (12)$$

where $\alpha$ is the Hilbert damping parameter. Substituting Eq. (2) into Eq. (12), and neglecting the terms containing $\varepsilon$ and $\beta$ in powers higher than 1st, one can calculate the derivatives of $\mathcal{R}$ with respect to generalized velocities $\dot{\varphi}$ and $\dot{\theta}$. Lagrange equations for angles $\varphi$ and $\theta$ can be obtained using Eq.(8) as:

$$-\frac{\mathcal{M}^2 \ddot{\theta}}{\delta \bar{\gamma}^2} + \frac{\mathcal{M}^2}{\delta}\frac{H}{\bar{\gamma}}\dot{\varphi}\cos\varphi - \frac{\mathcal{M}^2}{\delta}\left(\frac{\dot{\varphi}^2}{\gamma^2}\sin\theta\cos\theta - H^2\sin\theta\cos\theta\cos^2\varphi +\right.$$

$$\left. -H\cos\varphi\cos 2\theta\frac{\dot{\varphi}}{\bar{\gamma}}\right) - \dot{\varphi}\frac{m}{\gamma}\cos\theta - mH\sin\theta\cos\varphi + K\sin 2\theta = \frac{\alpha \mathcal{M}}{\gamma}\dot{\theta} \quad (13)$$

$$\frac{\mathcal{M}^2}{\delta\gamma}\left(-\frac{\dot{\varphi}}{\gamma}\dot{\theta}\sin 2\theta + H\dot{\theta}\cos 2\theta\cos\varphi - H\dot{\varphi}\sin\theta\cos\theta\sin\varphi + \frac{\ddot{\varphi}}{\gamma}\cos^2\theta\right) - \frac{m}{\gamma}\dot{\theta}\cos\theta +$$

$$+\frac{\mathcal{M}^2}{\delta}\left(\frac{\dot{\varphi}}{\gamma}H\sin\theta\cos\theta\sin\varphi + H^2\sin^2\theta\cos\varphi\sin\varphi\right) + \frac{\mathcal{M}^2}{\delta}\left(\frac{\dot{\theta}}{\gamma} - H\sin\varphi\right)H\cos\varphi \quad . \quad (14)$$

$$+ mH\cos\theta\sin\varphi = -\frac{\alpha\mathcal{M}}{\gamma}\dot{\varphi}\cos^2\theta$$

The further analysis of the behavior of the angles of the antiferromagnetic vector and the calculation of the frequencies of the corresponding oscillations require the linearization of equations (13)-(14) near the equilibrium states considered in the previous section. Substituting $\varphi=\varphi_0+\varphi_l$ and $\theta=\theta_0+\theta_l$, where $\varphi_0$ and $\theta_0$ correspond to the equilibrium angular coordinates of the vector **L**, $\varphi_l\ll 1$ and $\theta_l\ll 1$ are small deviations, so that higher than the first orders of these values are neglected, one can obtain the following equations. For the non-collinear phase ($\theta_0\neq 0$):

$$\ddot{\theta}_l + \frac{\alpha\mathcal{M}\gamma}{\chi_\perp}\dot{\theta}_l + \left(\gamma^2 H^2 + \frac{2K\gamma^2}{\chi_\perp}\right)(1-\cos^2\theta_0)\theta_l \mp \left(2\bar{\gamma}H\cos^2\theta_0 - \frac{m\gamma}{\chi_\perp}\cos\theta_0\right)\dot{\varphi}_l = 0$$
$$\ddot{\varphi}_l\cos^2\theta_0 + \frac{\alpha\mathcal{M}\gamma}{\chi_\perp}\dot{\varphi}_l\cos^2\theta_0 + \cos^2\theta_0\left(\frac{2K\gamma^2}{\chi_\perp}\right)\varphi_l \pm \left(2\bar{\gamma}H\cos^2\theta_0 - \frac{m\gamma}{\chi_\perp}\cos\theta_0\right)\dot{\theta}_l = 0$$
(15)

For the collinear phase ($\theta_0=0$):

$$\ddot{\theta}_l + \frac{\alpha\mathcal{M}\gamma}{\chi_\perp}\dot{\theta}_l + \left(\frac{|m|\gamma^2}{\chi_\perp}H - \gamma^2 H^2 - \frac{2K\bar{\gamma}^2}{\chi_\perp}\right)\theta_l \mp \left(2\gamma H - \frac{m\gamma}{\chi_\perp}\right)\dot{\varphi}_l = 0$$
$$\ddot{\varphi}_l + \frac{\alpha\mathcal{M}\gamma}{\chi_\perp}\dot{\varphi}_l + \left(\frac{|m|\gamma^2}{\chi_\perp}H - \gamma^2 H^2\right)\varphi_l \pm \left(2\gamma H - \frac{m\gamma}{\chi_\perp}\right)\dot{\theta}_l = 0$$
(16)

The first set of signs in each equation corresponds to the case $m>0$ $\theta_0>1$ or $m<0$ $\theta_0<1$, while the second one to the to the other cases.

Eqs. (15) and (16) have harmonic solutions $\theta_l=\theta_A\exp(i\omega t)$ and $\varphi_l=\varphi_A\exp(i\omega t)$ with the frequencies $\omega$ determined as the zeros of Eq. (15),(16) determinant:

$$\omega_{q-AFM,q-FM} = \left(\Omega_1^2 + \Omega_2^2 + \omega_0^2 \pm \sqrt{\left(\Omega_1^2+\Omega_2^2\right)^2 + 2\omega_0\Omega_2^2}\right)^{\frac{1}{2}} \quad (17)$$

for $0\leq |m| < \chi_\perp H + 2K/H$; and

$$\omega_{q-AFM,q-FM} = \left(\Omega_3^2 + \Omega_4^2 \pm \frac{1}{2}\sqrt{\omega_0^4 + 16\Omega_3^2\Omega_4^2}\right)^{\frac{1}{2}} \quad (18)$$

for $|m| \geq \chi_\perp H + 2K/H$.

In (17) and (18) $\omega_0 = \gamma\sqrt{2K/\chi_\perp}$, $\omega_{KK} = \gamma|m|/\chi_\perp$, $\omega_H = \gamma H$, $\Omega_1^2 = \left(\omega_H^2 - 2\omega_H\omega_{KK}\cos\theta_0\right)/2$, $\Omega_2^2 = \left(2\omega_H\cos\theta_0 - \omega_{KK}\right)^2/2$, $\Omega_3 = \omega_H - \omega_{KK}/2$, $\Omega_4 = \sqrt{\omega_K^2 - 2\omega_0^2}/2$. The choice of the sign before the root corresponds to the choice of the higher or lower frequency branch, the quasi-ferromagnetic and quasi-antiferromagnetic modes, respectively. The frequency relations are invariant to sign inversion of *m*. Notice that in case of significant difference of the gyromagnetic ratios of the sublattices, e.g. $\gamma_R \neq \gamma_{Fe}$, the frequencies of quasi-FM and quasi-AFM modes differ below and above compensation point $\omega_j(m) \neq \omega_j(-m)$, while the equilibrium states $\theta_0(m) = \theta_0(-m)$ are the same. Detailed consideration of these cases is out of the scope of the present study.

Eq. (18) remains valid for the magnetization compensation point *m*=0 and $\cos\theta_0$=0:

$$\omega_{q-AFM,q-FM} = \sqrt{\omega_0^2 + \omega_H^2/2 \pm \omega_H^2/2} \qquad (19)$$

and can be obtained directly from the linearized Lagrange equations for the other coordinate system choice with Y axis directed along the easy axis of the effective anisotropy **n** which helps to avoid uncertainty of $\varphi$ for *m*=0, and to avoid consequent zeroing of the precession equation, unlike Eq. (16).

[1]   S. Geller, J. P. Remeika, R. C. Sherwood, H. J. Williams, and G. P. Espinosa, *Magnetic Study of the Heavier Rare-Earth Iron Garnets*, Physical Review **137**, A1034 (1965).

[2]   M. S. Lataifeh and A. Al-sharif, *Magnetization Measurements on Some Rare-Earth Iron Garnets*, Appl Phys A Mater Sci Process **61**, 415 (1995).

[3]   Zvezdin A. K., *Dynamics of Domain Walls in Weak Ferromagnets*, ZhETF Pisma Redaktsiiu **29**, 605 (1979).

[4]   M. D. Davydova, K. A. Zvezdin, A. v Kimel, and A. K. Zvezdin, *Ultrafast Spin Dynamics in Ferrimagnets with Compensation Point*, Journal of Physics: Condensed Matter **32**, 01LT01 (2020).

[5]   T. G. H. Blank, K. A. Grishunin, E. A. Mashkovich, M. v. Logunov, A. K. Zvezdin, and A. v. Kimel, *THz-Scale Field-Induced Spin Dynamics in Ferrimagnetic Iron Garnets*, Phys Rev Lett **127**, 37203 (2021).